\documentclass[aip,jap,reprint,raggedbottom]{revtex4-2}
\usepackage{graphicx}
\usepackage{hyperref}
\usepackage{microtype}
\usepackage{xcolor}
\hypersetup{hidelinks,pdftitle={AgenticMM: A Material-Centered Multi-Agent Architecture for Scalable First-Principles Materials Modeling},pdfsubject={Material-centered multi-agent execution of structured prompt-based skills},pdfkeywords={AgenticMM, agentic AI, multi-agent systems, first-principles materials modeling, structured prompts, scientific workflows, DefectSkills}}

\begin{document}

\title{AgenticMM: A Material-Centered Multi-Agent Architecture for Scalable First-Principles Materials Modeling}
\author{Zhenkun Yuan}
\email{zkyuans@gmail.com}
\affiliation{Department of Materials Science and NanoEngineering, Rice University, Houston, TX 77005, USA}
\affiliation{Rice Advanced Materials Institute, Rice University, Houston, TX 77005, USA}
\author{Geoffroy Hautier}
\email{gh55@rice.edu}
\affiliation{Department of Materials Science and NanoEngineering, Rice University, Houston, TX 77005, USA}
\affiliation{Rice Advanced Materials Institute, Rice University, Houston, TX 77005, USA}
\date{\today}

\begin{abstract}
Multi-agent frameworks are attracting increasing attention as a means of autonomously driving complex, long-running workflows in first-principles materials modeling. However, such frameworks commonly organize AI agents by computational task or scientific role. When scaled across many materials, this organization can require repeated transfers of material-specific context and complicate persistent execution and resource coordination. Here, we introduce AgenticMM, a material-centered multi-agent architecture in which each material has a persistent agent that owns its complete workflow and retains its scientific context, while a persistent coordinator allocates shared resources, mediates communication, and tracks campaign progress. AgenticMM is designed to incorporate reusable, prompt-based skills as modular definitions of scientific workflows. Using DefectSkills prompts, we demonstrate the feasibility of AgenticMM through a multistage first-principles defect modeling campaign for five chemically diverse semiconductors, with five material-specific agents working concurrently, each relaxing the unit cell and performing host electronic-structure, phase-stability, and intrinsic-defect calculations for its assigned material.
\end{abstract}

\maketitle
\setcounter{dbltopnumber}{1}
\setlength{\abovecaptionskip}{5pt}
\setlength{\dbltextfloatsep}{13pt plus 2pt minus 2pt}

\section{Introduction}

AI agents based on large language models (LLMs) are enabling autonomous first-principles materials modeling by interpreting scientific objectives, executing computational workflows, and reasoning from intermediate results to decide what to investigate next,\cite{masgent,matclaw} capabilities that extend conventional workflow automation, as exemplified by atomate and AiiDA.\cite{atomate,aiida} For complex, long-running workflows, collaboration among multiple agents offers a promising route to scaling this autonomy and has motivated the development of multi-agent density functional theory (DFT) frameworks.\cite{dreams,autodft,tritondft}

Existing multi-agent frameworks commonly organize AI agents by computational task or scientific role, such as planning, input generation, simulation execution, monitoring, and analysis.\cite{dreams,autodft,tritondft,matsciagent,vriza2026} This organization can be effective for a complex study of a single material. In a campaign spanning many materials, however, agents may need to pass along each material's scientific context as responsibility shifts from one specialist to the next. These transfers can increase coordination overhead and introduce data gaps or referential drift.\cite{cemri2025,agentask2026} Such errors could lead to inconsistent decisions or cause an agent to use information from the wrong material.

To address these challenges, we introduce AgenticMM, a material-centered multi-agent architecture for scalable first-principles materials modeling. For each material, AgenticMM assigns one persistent agent that owns the complete workflow and retains material-specific context and execution history. A persistent coordinator agent allocates shared resources across dependency-ready tasks, mediates communication, routes unresolved scientific questions to the researcher, and tracks campaign-level completion. AgenticMM is designed to incorporate reusable, prompt-based skills as modular definitions of scientific workflows.\cite{atomisticskills,defectskills} We refer to these supplied skills as MatterSkills.

This paper proceeds as follows. Section~II presents the architecture and its design principles, Section~III defines mechanisms for persistent autonomous execution, and Section~IV demonstrates AgenticMM through a first-principles defect modeling campaign in which five material agents concurrently apply DefectSkills.\cite{defectskills} The campaign covers trigonal selenium (t-Se), Sb$_2$Se$_3$, CaMg$_2$P$_2$, CaZn$_2$N$_2$, and BaCu$_2$GeS$_4$. These materials include emerging or candidate solar absorbers\cite{moustafa2024,kavanagh2025,wang2024,pike2025,hinuma2016,fabini2019,vermang2026} and were selected to span elemental, binary, ternary, and quaternary compositions, providing a chemically diverse test set.

\section{Architecture and design principles}

\begin{figure*}[t]
  \centering
  \includegraphics[width=0.92\textwidth]{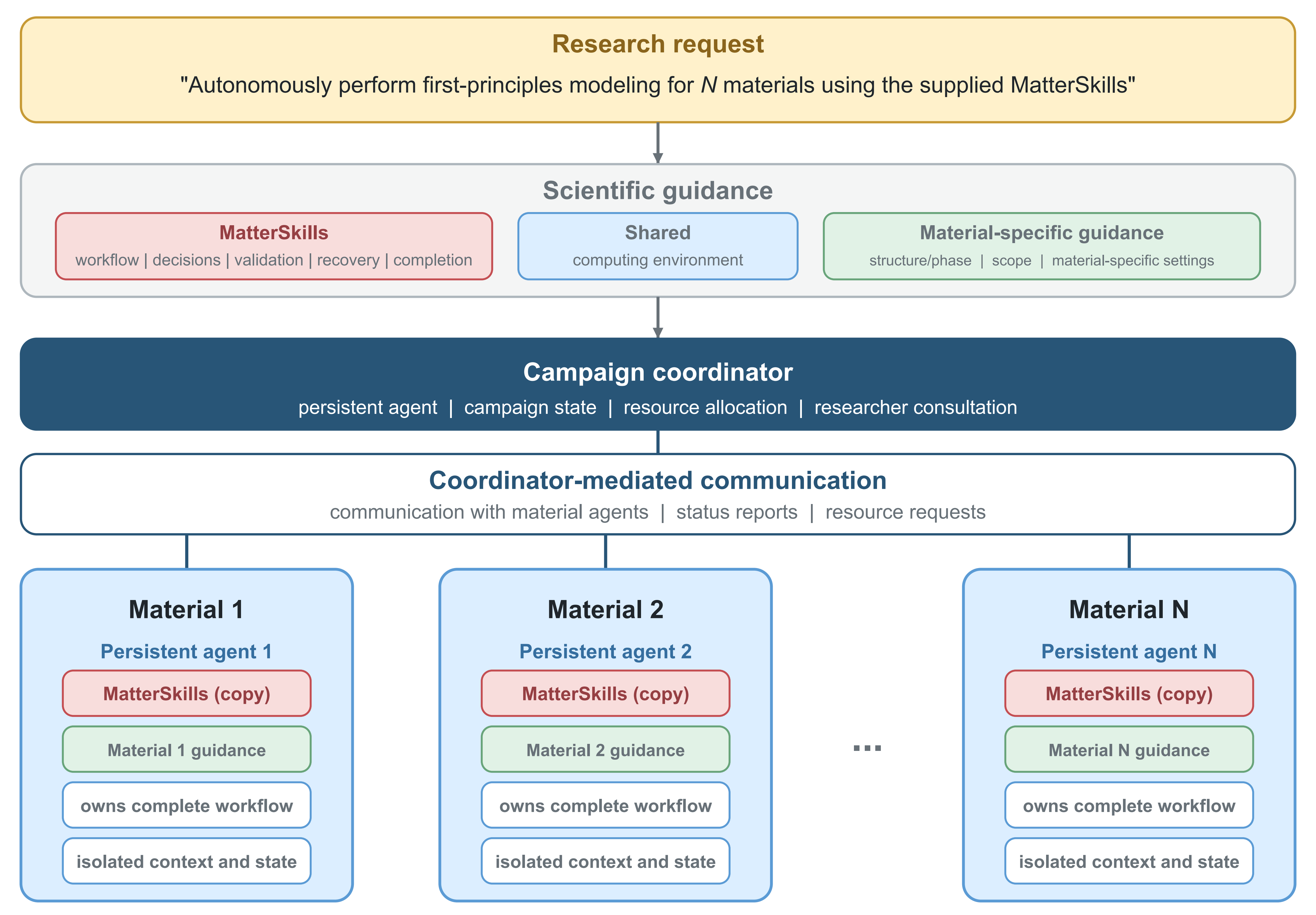}
  \caption{Architecture of AgenticMM. Supplied MatterSkills provide the primary scientific guidance, while researcher inputs define the campaign objective, shared computing environment, and scope and requirements for each material. A persistent coordinator agent distributes the applicable guidance, manages campaign state and shared resources, and mediates communication. Each persistent material agent receives the pinned MatterSkills release and relevant material-specific guidance, owns the complete MatterSkills-defined workflow, and maintains isolated context and state for one material. Material agents communicate only with the coordinator and do not share material-specific information with one another.}
  \label{fig:agenticmm_architecture}
\end{figure*}

As shown in Fig.~\ref{fig:agenticmm_architecture}, AgenticMM organizes the campaign around material-centered ownership. Throughout the campaign, each material agent owns its material's workflow and scientific context, while the coordinator manages campaign-wide information, shared resources, and communication. A campaign involving $N$ materials has exactly one coordinator role and one material-agent role per material, with no additional agent roles or layers. Supplied MatterSkills and researcher inputs provide the scientific guidance for the campaign. The design principles are detailed below.

\noindent\textbf{Material-centered ownership.} For each material, one persistent agent is the sole owner of the complete MatterSkills-defined workflow and its authoritative material-specific state, including inputs, calculation settings, intermediate results, decisions, and records of execution, failures, interruptions, and recovery actions. Material-specific information is not shared with other material agents. This ownership boundary reduces the risk of cross-material context mixing and allows materials to progress independently and at different rates.

\noindent\textbf{Campaign-level coordination.} The campaign coordinator agent maintains campaign-level state and coordinates activities across materials. It records the campaign objective, material-agent assignments and reassignments, shared settings and decisions, compact status reports, resource allocations, and completion state. During campaign setup, it assigns each material to one persistent agent and distributes the pinned MatterSkills and relevant researcher inputs, including the shared computing environment and the scientific scope and requirements for that material. During execution, it allocates shared resources based on material agents' reports of task readiness and resource needs, tracks progress toward campaign completion, and routes unresolved scientific questions to the researcher. Material agents communicate only with the coordinator agent and do not exchange material-specific information with one another. The coordinator agent neither performs material-level calculations nor aggregates detailed scientific state across materials. The coordinator and material-agent roles are persistent logical roles whose authoritative states survive interruptions and enable resumption or reassignment. Persistence does not require indefinitely running conversations.

\noindent\textbf{Separation between scientific workflows and the multi-agent architecture.} Supplied MatterSkills define the scientific workflow through instructions, decision rules, validation criteria, recovery procedures, and completion criteria. During campaign setup, researcher inputs define the overall campaign objective, shared computing environment, and the scientific scope and requirements for each material. This guidance is supplied independently of AgenticMM. The architecture organizes agents, communication, resources, and persistent execution without modifying the scientific workflow. If the supplied guidance does not resolve a scientific question arising during the campaign, the resulting researcher decision is recorded as supplemental guidance. All material agents use the same pinned version of the supplied MatterSkills together with the relevant recorded researcher decisions, supporting consistent execution and reproducibility.

These design principles make AgenticMM a general-purpose architecture in which each persistent material agent owns an isolated workflow and state, the coordinator manages campaign-wide communication and shared resources as materials progress independently, and supplied scientific workflows serve as modular inputs to multi-agent coordination and execution.

\section{Mechanisms for persistent autonomous execution}

This section specifies the execution protocol for AgenticMM. The campaign in Sec.~IV exercised some of these mechanisms and informed the design of others. Together with the campaign objective, researcher inputs, and the pinned MatterSkills release, these mechanisms are designed to enable a campaign to reach validated completion despite failures, session interruptions, or agent replacement while keeping shared resources productively occupied whenever a ready job can use them.

\noindent\textbf{Execution readiness and enforcement.} While AgenticMM defines agent roles and execution rules, it is not a standalone software system. Before launch, the coordinator verifies that the supplied MatterSkills and researcher inputs jointly specify the required workflow stages, inputs, dependencies, outputs, validation and completion criteria, permitted recovery actions, and conditions for researcher consultation. It also confirms that the runtime can instantiate the required roles and that the required data, approved computing environment, submission interface, scheduler, and enforcement utilities are accessible. All execution rules must be checkable, their application must be recordable, and execution state must be externalized. Deterministic safeguards, including task identifiers, input hashes, directory checks, submission locks, and scheduler reconciliation, must be enforced by tested, version-pinned utilities. The coordinator checks only operational completeness and enforceability. It must never judge or alter scientific content. If an instruction required for execution is missing, internally contradictory, or unenforceable, the coordinator identifies the gap, warns the researcher, and does not launch the campaign.

\noindent\textbf{Researcher-confirmed setup.} Before authorizing any execution, including the creation of calculation files, the coordinator asks any outstanding campaign-wide and material-specific questions one at a time, records the answers in the appropriate manifests, and obtains explicit researcher approval of the resulting setup.

\noindent\textbf{Role instantiation and ownership.} After researcher approval, the coordinator confirms its role and uses the runtime to instantiate one persistent agent per material. The coordinator records the assignments. Before execution, it verifies that the number of instantiated material agents matches the number of materials and that each material has exactly one owner. It reports the result. Directories or workflow tracks alone are insufficient. Each material agent acknowledges its assignment and demonstrates access to its workspace, guidance, and execution tools. The coordinator records these checks. Before each material-level action, the enforcement utility verifies that the acting agent owns the material. This applies to structure retrieval, file preparation or modification, job submission and monitoring, recovery, and analysis. The coordinator performs no material-level work. If an owner becomes unavailable or loses required access, the coordinator pauses work for that material until the owner can resume or a replacement handoff is recorded.

\noindent\textbf{Scientific authority.} The supplied MatterSkills remain the primary scientific authority throughout the campaign. At startup or after resumption or reassignment, every agent reloads the campaign objective, its role-specific guidance, and the pinned MatterSkills release. Before each major workflow stage, scientific decision, or recovery action, a material agent consults the relevant MatterSkills instructions and recorded researcher decisions. If the MatterSkills instructions and recorded researcher decisions do not resolve a scientific question, the material agent records it, notifies the coordinator, and continues all work that does not depend on its resolution. The coordinator routes the question to the researcher, records the researcher-approved resolution, and returns it to the material agent before dependent work resumes.

\noindent\textbf{Dependency-aware resource use.} Each material agent derives and maintains a dependency graph from the pinned MatterSkills release and its confirmed manifest. It marks a task ready only after its prerequisites are satisfied and any required upstream results have passed validation. It then reports every ready task and its resource requirements without transferring the full material-specific graph. The coordinator allocates resources among these tasks without imposing a one-job-per-material limit, allowing independent tasks for the same material to run concurrently. All attempts associated with a task, including its initial submission, retries, and continuations, share a task-level lock. Material-wide locks are reserved for operations that modify shared material state. When a branch cannot proceed, the coordinator reallocates available resources to compatible ready tasks from another branch or material whenever possible.

\noindent\textbf{Validated submission and monitoring.} A material agent validates the calculation inputs and requests resources using a stable task identifier. The task identifier persists across retries and continuations, and each submission attempt receives a distinct attempt identifier. Before any recovery action, the agent records the applicable guidance rule, authorized action, and target directory. The coordinator authorizes resources based on reported readiness and available capacity without changing the calculation inputs or recovery action. After authorization, the material agent invokes the shared submission interface, which applies the required checks. If a check fails, the interface blocks submission and records the diagnostic. The material agent reports the blocked submission to the coordinator so it can release any reserved capacity. After successful submission, the material agent records the scheduler identifier and reports the task identifier, attempt identifier, scheduler identifier, and allocated resources to the coordinator. It then monitors the job and reports its completion or failure and the released resources. Scheduler completion is recorded separately from scientific validation because a successful process exit does not establish a scientifically valid result. After a task fails, only guidance-permitted recovery actions may proceed for that task.

\noindent\textbf{Persistent state and recovery.} Authoritative material state includes the pinned MatterSkills release, material manifest, ledger, event log, calculation files, and scheduler records. Authoritative campaign state includes the campaign manifest, ledger, event log, resource and scheduler records, and compact material-agent reports. Manifests record confirmed setup, ledgers track current execution state, and append-only event logs record transitions. Agent activity is logged separately from scheduler and calculation activity. Related systems similarly use persistent project records or structured external state for restartability.\cite{catmaster,elagentegraphico} Before any state-changing external action, the responsible agent appends an intent event recording a stable event identifier, the planned action, and any applicable task and attempt identifiers, input and configuration hashes, working directory, and expected artifacts. Afterward, it appends a corresponding outcome event recording success or failure, any applicable scheduler identifier, and the resulting state. Before updating records or issuing an external action, a resuming or newly assigned agent obtains exclusive ownership of its role and verifies that no previous instance can act. It then reloads its authoritative state and replays only its role's event log. A material agent reconciles its ledger with calculation files and scheduler records. The coordinator reconciles the campaign ledger with compact material-agent reports and campaign-level resource and scheduler records. Any conflict must be reported before further action. The resuming or newly assigned agent must not duplicate active work or validated artifacts.

\noindent\textbf{Completion and selective reopening.} A material agent reconciles its dependency graph using stable task identifiers, not filenames or job counts. It declares the material complete only after every required task passes the validation checks in the applicable guidance and the material passes the final reliability assessment. Researcher-authorized follow-up calculations reopen only the affected portion of the material workflow and invalidate dependent results as required by the supplied guidance. The coordinator declares the campaign complete only after every material meets its recorded completion criteria.

\begin{figure*}[t]
  \centering
  \includegraphics[width=\textwidth]{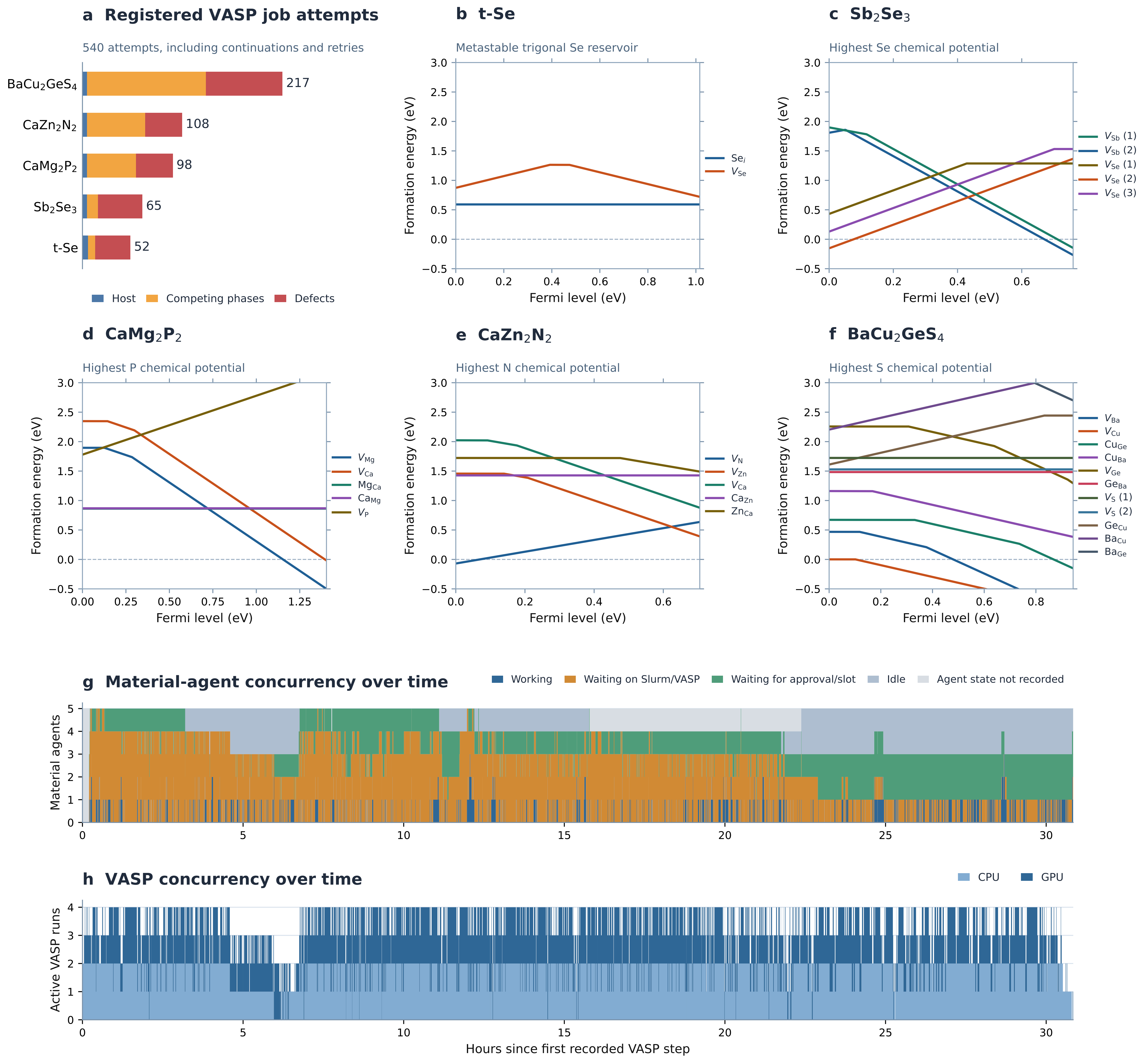}
  \caption{Snapshot of the completed clean-room five-material defect campaign. (a) 540 registered VASP job attempts for host, competing-phase, and defect calculations, including continuations and retries. (b)--(f) Representative defect formation-energy diagrams for t-Se, Sb$_2$Se$_3$, CaMg$_2$P$_2$, CaZn$_2$N$_2$, and BaCu$_2$GeS$_4$ under the chemical-potential conditions stated in each panel. The material agents reported formation energies for a total of 109 defect configurations. (g) Activity of the five material agents during the 30.8-h interval from the first to the last recorded VASP step. The thickness of each band shows how many agents were in the indicated state at each time. Light gray denotes intervals for which an agent's state was not recorded. No agent state was recorded in the first 12~min after VASP started. This and later gaps, mainly during Sb$_2$Se$_3$ postprocessing, account for about 5\% of total agent time in panel (g). (h) Concurrent CPU and GPU VASP runs over the same interval.}
  \label{fig:defect_campaign_results}
\end{figure*}

\section{Defect calculations as a demonstration}

We demonstrate AgenticMM in a five-material point-defect modeling campaign using DefectSkills\cite{defectskills} as the supplied MatterSkills. DefectSkills comprises a set of structured prompts that define the workflow for VASP-based first-principles defect calculations in semiconductors and insulators, including scientific procedures, decision rules, validation criteria, and recovery procedures. This workflow spans unit-cell relaxation, phase stability and chemical potentials, electronic structure, dielectric response, charged-defect supercell generation and calculations, finite-size corrections, and defect thermodynamics. These branches differ in prerequisites, computational cost, convergence criteria, required outputs, and failure modes. We note that the purpose of this campaign is to evaluate AgenticMM. Reference-quality defect results would require careful convergence tests, configurational searches, and physical validation.\cite{freysoldt2014}

The campaign began in a clean-room session in the OpenAI Codex graphical user interface, with no inherited conversational memory. The coordinator and five material agents used GPT-6.1 Sol with high reasoning effort and relied on supplied documents, researcher-confirmed inputs, and campaign state recorded during the run. No role-specific model specialization was used, and AgenticMM does not depend on this model configuration.

At the start of the session, we supplied the then-current PDFs describing the AgenticMM architecture and execution protocol (\texttt{AgenticMM.pdf}) and the DefectSkills workflow (\texttt{DefectSkills.pdf}), and entered the following message in the Codex chat:
\begin{quote}
\itshape
I want to perform defect calculations for five materials, trigonal Se, Sb$_2$Se$_3$, CaZn$_2$N$_2$, CaMg$_2$P$_2$, and BaCu$_2$GeS$_4$, using the AgenticMM architecture and the DefectSkills.
\end{quote}
The coordinator recorded SHA-256 checksums identifying the supplied document versions. It connected to NERSC Perlmutter under a clean-room access restriction that prohibited reading any pre-existing file, directory, or calculation without explicit researcher permission. After the calculations had begun, we asked the coordinator to display the campaign goal. The following excerpt summarizes the approved scope and completion requirements.
\begin{quote}
\itshape
Complete the five-material intrinsic point-defect campaign on NERSC Perlmutter with one AgenticMM coordinator and five dedicated material agents. Follow the approved campaign transcript and DefectSkills for trigonal Se, Sb$_2$Se$_3$, CaZn$_2$N$_2$, CaMg$_2$P$_2$, and BaCu$_2$GeS$_4$. Validate each material's PBE calculations, defect formation energies and thermodynamics, and backup. Respect the approved access scope.
\end{quote}

During campaign setup, the coordinator instantiated five dedicated material agents as Codex subagent threads, assigned one exclusively to each material, and recorded the ownership assignments in the campaign manifest. The agents operated under shared computational limits, and the coordinator allocated resources as the heterogeneous workflows progressed asynchronously.

Under the supplied guidance, material agents performed setup, monitoring, and permitted recovery. The researcher refined the scientific scope, resolved questions outside DefectSkills, reconfirmed access permissions, and directed a change to the launcher's locking policy.

Calculations ran on NERSC Perlmutter with two concurrent interactive GPU allocation slots and two concurrent interactive CPU allocation slots, each spanning four nodes. All calculations used the projector-augmented-wave method as implemented in VASP\cite{paw} and the GGA-PBE functional.\cite{pbe} The pristine supercells used for defect calculations in t-Se, Sb$_2$Se$_3$, CaMg$_2$P$_2$, CaZn$_2$N$_2$, and BaCu$_2$GeS$_4$ contained 81, 60, 90, 90, and 96 atoms, respectively. The defect scope included vacancies, applicable cation--cation antisites, and Se interstitials in t-Se. Other interstitials and cation--anion antisites were deferred. The selected functional, supercell sizes, defect scope, and competing-phase cutoff of 20~meV/atom above the convex hull were the principal campaign-wide departures from DefectSkills defaults. Material-specific choices were recorded in the campaign manifests, and all other settings followed the supplied DefectSkills.

Figure~\ref{fig:defect_campaign_results} presents the campaign workload, representative scientific results, and concurrency metrics. Panel (a) summarizes 540 registered job attempts, including continuations and retries. Of these attempts, 535 reached VASP execution: 524 completed and 11 failed. The remaining five were canceled before VASP started. Of the 535 VASP runs, 297 ran on GPUs and 238 on CPUs. Panels (b)--(f) show representative defect formation-energy results.

The demonstration evaluated context isolation, concurrent activity among material agents, use of computing slots whenever a ready job matched the available resources, prevention of duplicate work during recovery, and completion under DefectSkills and the researcher-approved reporting policy.

We find that, throughout the campaign, each material agent operated within its assigned material directory and maintained a separate recorded state, with no observed cross-material context mixing. Figure~\ref{fig:defect_campaign_results}(g) shows overlapping periods of work among the material agents during the VASP interval. The campaign log records all five agents working simultaneously during final reporting. These findings demonstrate compliance with the ownership boundary during the campaign, although no quantitative context-bleed test was performed.

Panel (h) traces concurrent CPU and GPU VASP runs over the same 30.8-h interval. Active-slot occupancy was 82.9\% for the two GPU slots and 86.5\% for the two CPU slots. These values measure calculation-slot occupancy rather than hardware-level processor utilization. The main orchestration issue was an initial launcher that held an exclusive material-level lock and unnecessarily restricted concurrency. Replacing it with task-level locks allowed independent jobs from the same material agent to run simultaneously, and the BaCu$_2$GeS$_4$ agent subsequently used all four slots. The remaining gaps coincided with scheduler waits, recovery work, or intervals without a recorded compatible ready task. The accounting does not assign a unique cause to every gap. These findings motivated the task-level locking and dependency-aware resource management mechanisms in Sec.~III.

Of the 11 VASP runs that Slurm marked as failed, four encountered eigensolver errors and seven encountered ionic line-search failures. Scientific validation also detected electronic nonconvergence in an Sb$_2$Se$_3$ calculation that exited successfully. All affected calculations, including those associated with five pre-VASP cancellations, subsequently completed through guidance-permitted retries or continuations, leaving no failed calculations at campaign completion. The agents archived each failed attempt, verified restart inputs or saved geometries, obtained validated VASP outputs, and avoided duplicate active submissions. Some recoveries required researcher decisions or corrections to execution utilities, so an autonomous recovery rate was not quantified. These findings informed the mechanisms for scientific authority, validated submission, and persistent state specified in Sec.~III.

The VASP outputs passed calculation-level validation. The researcher then authorized reporting all 109 defect configurations without final postprocessing validation, with unresolved charge corrections and band-edge choices documented. The campaign was marked complete under the researcher-approved reporting policy after final reconciliation of campaign state, material activity logs, the job registry, incident records, and checksum-verified backup manifests. Two mechanisms specified in Sec.~III were not fully evaluated: the intent--outcome event scheme was not applied to every state-changing external action, and the execution-readiness assessment was not performed before campaign launch. No agent required replacement during the campaign, so recovery by a newly assigned agent was not tested. These mechanisms and the replacement-agent recovery scenario remain to be validated through controlled tests in future deployments.

The five-material campaign demonstrates the feasibility of material-centered execution for multistage defect calculations. Applying AgenticMM to other first-principles materials-modeling domains requires MatterSkills that define the workflow and dependencies, calculation-directory hierarchy, validation and recovery rules, consultation conditions, and completion criteria. Future demonstrations using different or combined MatterSkills, together with comparisons to task-centered agent organizations, can test broader applicability and identify the classes of problems for which material-centered ownership is most advantageous.

\section{Conclusion}

In summary, we present AgenticMM, a general-purpose, material-centered multi-agent architecture for scalable first-principles materials modeling. In AgenticMM, each material is assigned to a persistent agent, a coordinator manages shared resources and campaign-wide communication, and externally supplied MatterSkills define the scientific workflows. Using DefectSkills, we tested AgenticMM in a clean-room five-material defect campaign. The campaign maintained isolated material states, ran five material agents concurrently, coordinated shared CPU and GPU resources, and recovered all calculations affected by failed or canceled attempts without duplicate active submissions. It also identified practical issues including material-level locks that blocked independent calculations, temporary gaps in VASP concurrency, and recoveries requiring researcher decisions or corrections to execution utilities. These observations shaped the execution mechanisms proposed here. Complete intent--outcome logging, prospective readiness assessment, and replacement-agent recovery still require controlled validation. AgenticMM thus offers a material-centered approach to distributing autonomous first-principles workflows across many materials and scaling toward high-throughput execution. A future extension could let material agents delegate specific tasks to subagents while retaining ownership of their workflows, scientific context, and execution state.

\section*{Declaration of AI use}

OpenAI Codex was used to assist with developing and refining AgenticMM, coordinating and executing the five-material demonstration, preparing computational inputs, monitoring and recovering calculations, analyzing outputs, preparing figures, and editing the manuscript. The campaign coordinator and all five material agents used the GPT-6.1 Sol model with high reasoning effort in the OpenAI Codex graphical user interface. The authors defined the research scope and computational requirements, reviewed the AI-assisted work, and assessed the reported results and their limitations. The authors take full responsibility for the scientific content and conclusions.

\section*{Acknowledgments}

This work was supported by the U.S. Department of Energy, Office of Science, Basic Energy Sciences in Quantum Information Science under Award No.~DE-SC0022289. This research used resources of the National Energy Research Scientific Computing Center, a DOE Office of Science User Facility supported by the Office of Science of the U.S. Department of Energy under Contract No.~DE-AC02-05CH11231, using NERSC award BES-ERCAP0020966. Zhenkun Yuan acknowledges helpful discussions with Chenmu Zhang (Rice University).

\section*{Conflict of interest}

The authors have no conflicts of interest to disclose.

\section*{Data availability}

The AgenticMM prompts are available at \url{https://github.com/zy75-prog/AgenticMM}. Supporting campaign records and analysis and figure-generation scripts will be released in a public repository upon publication. Supporting calculation data will be shared subject to VASP licensing restrictions.

\end{document}